\documentclass[aps,twocolumn,superscriptaddress,groupedaddress,showkeys,showpacs,floatfix]{revtex4-1}
\usepackage{bm,amsmath,amssymb,dcolumn,siunitx,graphicx,float}         
\usepackage[version=4]{mhchem}

\begin{document}

\author{Ariana Torres-Knoop} 
	\affiliation{SURFsara, Science Park 140, 1098 XG  Amsterdam, the Netherlands}

\author{Ivan Kryven} 
	\email{i.kryven@uu.nl}
	\affiliation{Mathematical Institute, Utrecht University, Budapestlaan 6, 3508 TA Utrecht, the Netherlands}
	\affiliation{Centre for Complex Systems Studies, 3584 CE Utrecht, the Netherlands}

\title{Learning heterogenous reaction rates from stochastic simulations}

\begin{abstract}
Reaction rate equations are ordinary differential equations that are frequently used to describe deterministic chemical kinetics at the macroscopic scale.
At the microscopic scale, the chemical kinetics is stochastic and can be captured by complex dynamical systems reproducing spatial movements of molecules and their collisions. Such molecular dynamics systems may implicitly capture intricate phenomena that affect reaction rates but are not accounted for in the  macroscopic models.
In this work we present a data assimilation procedure for learning non-homogenous kinetic parameters from molecular simulations with many simultaneously reacting species. The learned parameters can then be plugged into the deterministic reaction rate equations to predict long time evolution of the macroscopic system. In this way, our procedure discovers an effective differential equation for reaction kinetics.
   To demonstrate the procedure, we upscale the kinetics of a molecular system that forms a complex covalently bonded network severely interfering with the reaction rates. Incidentally, we report that the kinetic parameters of this system feature a peculiar time and temperature dependences, whereas the probability of a network strand to close a cycle follows a universal distribution.
\end{abstract}

\maketitle 
\begin{figure*}[htbp]
\begin{center}
\includegraphics[width=\textwidth]{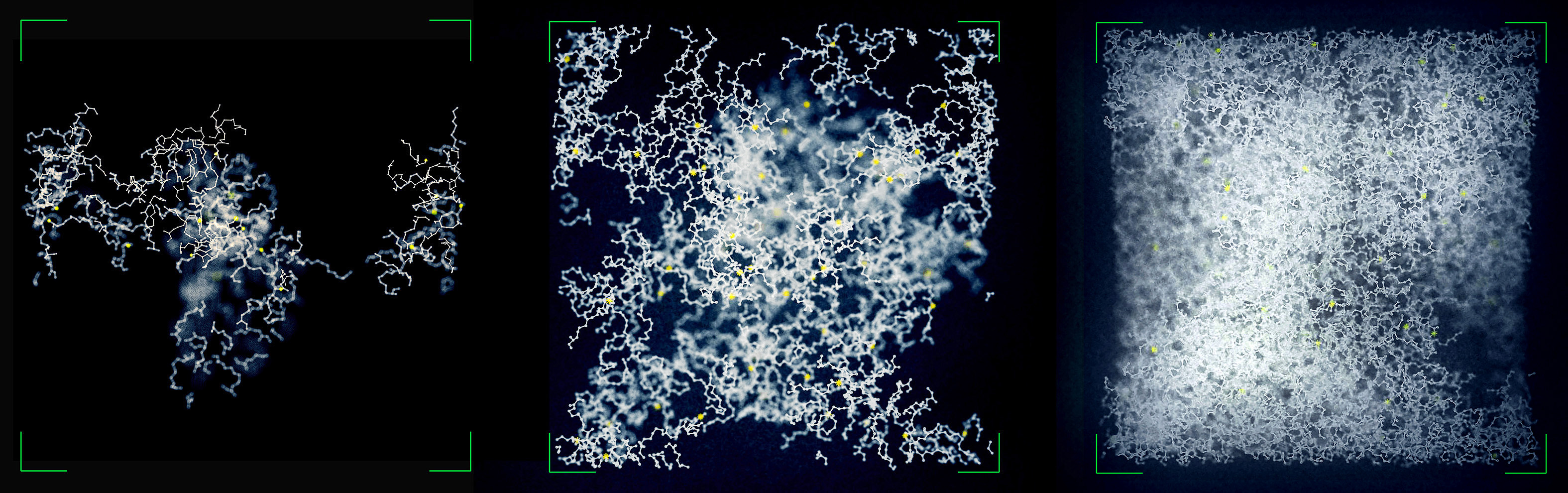}
\caption{(Colour online) Time snapshots of the carbon skeleton  of the largest cluster in the di-acrylate network as given by molecular  simulations suggest that the reaction rates may considerably slow down during the course of polymerisation. Left-to-right: 20\%, 30\%, and 80\% of reaction progress as measured by the double bond conversion, $\chi$.  }
\label{fig:boxes}
\end{center}
\end{figure*}

 \section{Introduction}
How to deduce chemical rate constants from observations? 
On the macroscopic scale, where concentrations of chemical compounds are deterministic quantities, this question was answered by Arrhenius who linked the reaction rate constants with slopes and intersection points of the concentration related profiles. 
Microscopic systems, as for instance, living cells \cite{gardner2000,kryven2015}, micropores \cite{branciamore2009}, or those used for \emph{in silico} computer experiments \cite{matsumoto2002,farah2012,omar2017,torres2018,torres2021}, typically have a small reaction volume, and therefore, the corresponding reaction rates may feature stochastic fluctuations that are not accounted for in the Arrhenius theory. 
Other assumptions of the Arrhenius theory, as the well-mixed environment, Boltzmann's stosszahlansatz, absence of memory, and non-cooperation of particles may lead to artefacts even in the case of macroscopic systems.
If such artefacts occur \cite{Ahmad2017,scolari2018,Simone2018}, the reaction rate constants appear time-dependent.
For example,  irreversible polymerisation leads to progressively growing molecules and therefore each reaction firing changes the conditions of the system, and consequently, the reaction rates  \cite{decker1996,rooney2018}. 
Molecular networks pose an especially severe case: their physical properties evolve considerably in the course of the assembly process and the latter may  undergo various types of phase transitions \cite{torres2018,omar2017,torres2021}. As an illustration of how strong such changes can be, Figure~\ref{fig:boxes} depicts formation of a percolating molecular network that significantly limits the mobility of all species. 

Molecular dynamics (MD) simulations \cite{farah2012} describe the evolution of a complex system by solving the equation of motion for each molecule and do not require reaction rate constants as input. For the purpose of this paper, we view the outcome of such simulations as large streams of data that implicitly contain information about the rates. Provided the reaction rates are extracted from these time series, the rates may be used as input for large-scale models, hence enabling a multi-scale paradigm. Among such macroscopic models are ordinary differential equations for species concentrations, chemical master equation,  Langevin equation, the Stochastic Simulation Algorithm (SSA) and other Monte Carlo methods \cite{higham2008}.

While the foundation of reaction rates is frequently discussed in the literature \cite{buff1960,weiss1986,hanggi1990,bolhuis2018}, this paper takes a phenomenological view and develops a practical method for inferring reaction rate parameters from noisy microscopic observations as given by, for example, molecular dynamics simulations.

 \section{Chemical rate equation}
Consider a system that consists of $N$ chemical species reacting via $M$ reactions. Each species may be represented by multiple particles, which is indicated by particle count vector $\boldsymbol x =(x_1,x_2,\dots,x_N)^\top$, where $x_i$ are the  numbers of copies that species $i$ is represented with. We thus have $\sum_{i=1}^N x_i$ particles in total. The reactive interactions that occur between these species can be modelled using three levels of mathematical description \cite{higham2008}: the equation of motion, stochastic process, and rate equation.  

The rate equations are ordinary differential equations (ODEs) that instead of species counts $x_i$, govern the evolution of their the molar concentrations $\boldsymbol c =(c_1,c_2,\dots,c_N)^\top$ with:
 \begin{equation}\label{eq:ctox}
 c_i = \frac{x_i}{ V N_A },
 \end{equation}
  whereby the volume $V\to\infty$ and $x_i$ are assumed to scale in such a way that keeps the pressure constant, and $N_A$ is the Avogadro's constant. In the general case of $M$ reactions, the ODEs are given by:
   \begin{equation}\label{eq:generalODE}
c_i'(t) =  \sum\limits_{j=1}^{M} k_j S_{i,j} \boldsymbol c^{\boldsymbol  \nu_j  } (t),\; i=1,2,\dots,N,
\end{equation}
where $k_j$ are the reaction rate constants,  $\boldsymbol \nu_{j}$ are binary vectors defining the participation of species $i$ in reaction $ j$,  and the vector power $\boldsymbol c^{\boldsymbol \nu}= c_1^{\nu_1} c_1^{\nu_2} \dots c_N^{\nu_N}$ is evaluated in the element-wise manner.  Matrix $\boldsymbol S$ has size $N\times M$ and is composed of stoichiometric vectors as its rows. For example, $S_{i,j}=1$ if  the $i^\text{th}$ species is the product of the $j^\text{th}$ reaction, $S_{i,j}=-1$ if it is a non-unique reactant, $S_{i,j}=-1$  if it is the only reactant in the second order reaction, and $S_{i,j}=0$ non-participant.

The intuition behind Eq. \eqref{eq:generalODE} becomes clearer after considering the following example. 
Consider a system that consists of three chemical species A,B and C, having the particle counts $\#\text{A}=x_1$, $\#\text{B}=x_2$, and $\#\text{C} = x_3$, and reacting via the following mechanism:
\begin{equation}\label{eqs:eg}
\ce{ A + B <=>[k_1][k_2]C }.
\end{equation}
By defining species concentrations with equation \eqref{eq:ctox}, we arrive with the following set of ODEs:
\begin{equation}\label{eqs:eg_ode}\begin{cases}
c_3' =  k_1 c_1c_2-k_2 c_3,&\\
c_2' =- k_1 c_1c_2+k_2 c_3,&\\
c_1' =- k_1 c_1 c_2+k_2 c_3,&\\
\end{cases}
\end{equation}
where $k_i$ are the rate constants.
In order to see that Eq. \eqref{eqs:eg_ode} is the special case of Eq. \eqref{eq:generalODE} it is sufficient to substitute:
$$
S=
\left(\begin{matrix}
-1 &-1& \;\;\,1\\
\;\;\,1 &- 1 &-1\\
\end{matrix}\right)^\top
,\; \boldsymbol \nu_1=(1, 1, 0)^\top,\; \boldsymbol \nu_2=(0, 0,1)^\top.
$$ 
One can see that the elements of $\boldsymbol \nu_1$ sum up to 2, which indicates that $j=1$ is a first order reaction, 
whereas the elements of $\boldsymbol \nu_2$ sum up to 1, indicating that the reaction order of $j=2$ is two.

\section{Stochastic rate equation}
We will now introduce a  stochastic rate equation that operates with discrete particle counts $x_i$ as opposed to  continuous concentrations used in \eqref{eq:generalODE}.   Suppose that all elements of the species count vector are large, $\boldsymbol x\gg0$, and in a small time increments $\tau$ these values undergo a small relative change. 
Let  $\boldsymbol z = (z_1,z_2,\dots,z_M)^\top$ be the column vector of reaction firings observed during  time interval $\tau$. We also assume  that the dynamics is a heterogenous renewal process,  that is the elements of $\boldsymbol z$  are independent Poisson random variables: $ z_j \sim \text{Poiss}[  \lambda_i\boldsymbol x^{\boldsymbol \nu_i} \tau],\; j=1,\dots,M,$ which, when combined with reaction stoichiometry $\boldsymbol S$,  provides the update vectors for species counts $\boldsymbol x$ at a given time interval. By iterating   $\tau_l=t_{l}-t_{l-1}$ over all discrete time intervals, one recovers the whole evolution trajectory of species count vector $\boldsymbol  x_{l}$ for $ l=1,\dots,L$: 
\begin{equation}\label{eq:walk}
\begin{aligned}
&\boldsymbol  x_{l} = \boldsymbol  x_{l-1} + \boldsymbol S \boldsymbol z_{l-1},\\
 & \boldsymbol z_l\sim (\text{Poiss}[  \lambda_1\boldsymbol x_l^{\boldsymbol \nu_1} \tau_l],\dots,\text{Poiss}[  \lambda_M\boldsymbol x_l^{\boldsymbol \nu_M}\tau_l])^\top,
  \end{aligned}
\end{equation}
where coefficients $\lambda_i$ are (time dependent) parameters related to  reaction rates $k_i$. Appendix I sketches the derivation of  equation \eqref{eq:walk} and explains the relationship between rate constants $k_i$ and $\lambda_i$.  This equation resembles an implementation of the $\tau$-leaping method \cite{gillespie2001} can be regarded as an $N$-dimensional random walk on species count numbers.

Stochastic process \eqref{eq:walk}, although practical, relies upon the system being well-mixed, memoryless, and non-cooperative among other assumptions. We suggest that this can be partially remedied for by inferring the time-dependent coefficients $\lambda_i$ from molecular simulations that does not suffer from these issues. 
 
\section{Data Assimilation Procedures}
In this section we assume that the empirical trajectories of the species counts $\boldsymbol {\tilde  x_{l}}$ and the counts of all reaction firings $\boldsymbol{\boldsymbol{\tilde z}_{l}}$ are known. We solve the inverse problem for 
estimating parameters $\lambda_1,\dots,\lambda_M$, which  may depend on time.
 Namely, we propose several statistical inference methods, so called maximum likelihood estimators (MLEs), for estimating effective reaction rates $\lambda_i$ that can be readily used in the stochastic model \eqref{eq:walk} or ODEs \eqref{eq:generalODE}. The source code implementing the estimators \eqref{eq:constMLE}-\eqref{eq:varMLE4} is provided  \footnote{https://github.com/ikryven/RateInference}.

\paragraph*{Constant Rate Estimator.} 
 Assuming that the stochastic rates $\lambda_j$  do not depend on time, then the following estimates hold:
\begin{equation}\label{eq:constMLE} 
\lambda_j=\frac{\langle{ \tilde  z}_{j,l}\rangle}{\langle \boldsymbol {\tilde x}_l^{\boldsymbol \nu_{j}}\tau_l\rangle},
\;
 \text{var}(\lambda_j) =  \frac{\lambda_{j}^2}{L\langle \boldsymbol{\tilde z}_{j,l}\rangle},
\end{equation}
where $$\langle x_l\rangle:=\frac{1}{L}\sum\limits_{l=1}^{L}x_l$$ denotes the time-average and $ \text{var}(\lambda_j)$ refers to the asymptotic variance of this estimator, which  may be used to derive the confidence intervals. 
See Appendix II  for the derivations.

\paragraph*{Moving-Average Rate Estimator.}
The following estimators yield rates in the form of a time series:
\begin{equation}\label{eq:seriesMLE}
\lambda_{j,l}=\frac{\langle{\tilde z}_{j,l}\rangle_s}{\langle\boldsymbol{\tilde x}_l^{\boldsymbol \nu_{j}}\tau_l\rangle_s},\;
\text{var}(\lambda_{l,j}) =  \frac{\lambda_{j,l}^2}{(2s+1)\langle {\tilde z}_{j,l}\rangle_s},
\end{equation}
where $$\langle x_l\rangle_s:=\frac{1}{2s+1}\sum\limits_{l=l-s}^{l+s}x_l$$ represents the moving average with window size $s$.
See Appendix III  for the derivations.

\paragraph*{Exponential Rate Estimator.}
 Consider the following ansatz  for the parameters of process \eqref{eq:walk}:
\begin{equation}\label{eq:MLE_EXPanz}
\lambda_{j}(t) = \lambda_{j,0}e^{-\alpha_j t}.
\end{equation}
 The estimators for the coefficients are given by:
$$
\lambda_{j,0} =  
\frac{\langle \boldsymbol{\tilde z}_{j,l}\rangle}{\langle e^{-\alpha_j t_l}\boldsymbol {\tilde x}_l^{\boldsymbol \nu_j}\tau_l\rangle}
$$
and
 $$
 \alpha_j =-\ln \omega_j,
  $$
  where $\omega_j \in [0,1]$  are the unique roots of
$
\langle \left( t_l\langle \boldsymbol{\tilde z}_{j,l}\rangle   -\langle\boldsymbol{\tilde z}_{j,l} t_l\rangle \right) \boldsymbol {\tilde x}_l^{\boldsymbol \nu_j}\tau_l \omega_j^{ t_l} \rangle =0
$ for $j=1,\dots,M.$
The variances of the exponents are given by:
   $$\text{var}(\alpha_{j}) =\frac{1}{L\lambda_{j,0}\langle t_l^2 e^{-\alpha_j t_l}\boldsymbol {\tilde x}_l^{\boldsymbol \nu_j} \tau_l\rangle}.$$
   and of the pre-factor by:
   $$\text{var}(\lambda_{j}) = \text{var}(\lambda_{j,0}) =  \frac{\lambda_{j,0}^2}{L\langle \boldsymbol{\tilde z}_{j,l}\rangle}.$$
  See Appendix IV  for the derivations.

\paragraph{Exp-Polynomial Rate Estimator.}
 Assume that the  reaction rate parameters that appear in the random walk model \eqref{eq:walk} have an exponential dependence on time of the form:
\begin{equation}\label{eq:MLE_EXP4}
\lambda_j(t)= e^{-p_j(t)},
\end{equation}
where   $p_j(t)=  \alpha_{j,0} +\alpha_{j,1}t+\alpha_{j,2}t^2+\dots+\alpha_{j,s}t^S,$
  is a polynomial of order $S$. For each $j$, the estimators of $\alpha_{j,s}$ are  found from the system of $S$ algebraic equations:
  \begin{equation}
\langle  (e^{-p_j(t_l)} \boldsymbol {\tilde x}_l^{\boldsymbol \nu_j}\tau_l - \boldsymbol{\tilde z}_{j,l}) t_l^s\rangle=0,\;s=0,\dots,S, \end{equation}
and the variances of rates' logarithms are  given by:
\begin{equation}\label{eq:varMLE4}
\text{var} ( \ln \lambda_j(t) )  =\frac{1}{L}  \boldsymbol b^{\top}\boldsymbol H^{-1}_j \boldsymbol b,
\end{equation}
where $\boldsymbol H_j$ are $(S+1)\times (S+1)$ matrices with elements:
$$(\boldsymbol H_j)_{k,s} = \langle  e^{-p_j(t_l)} t_l^{k} t_l^{s} \boldsymbol {\tilde x}_l^{\boldsymbol \nu_j} \tau_l\rangle.$$ 
and $\boldsymbol b = (1,t,t^2,\dots,t^S)^\top$. 
In fact, one can replace time $t$ in MLE  \eqref{eq:MLE_EXP4} with any monotonous function of time that tracks the progress of the chemical system, for example the conversion of an important species.  
The derivations are given in Appendix V.

\paragraph*{Model selection.}
There are two parameters describing the quality of the estimate that may be used when choosing the best MLE, and in the case of the polynomial estimator, when choosing the polynomial order.
 A small variance implies that the system is large enough to derive consistent estimates with a given estimator.
 A small residual implies that the estimator explains observed data. 
  In order to rationally determine the best order of the polynomial for approximation, we propose to minimise two qualities simultaneously: the variance and  residual.

\section{Example: Rates of Network Formation}

  \begin{figure}
\begin{center}
\includegraphics[width=1\columnwidth]{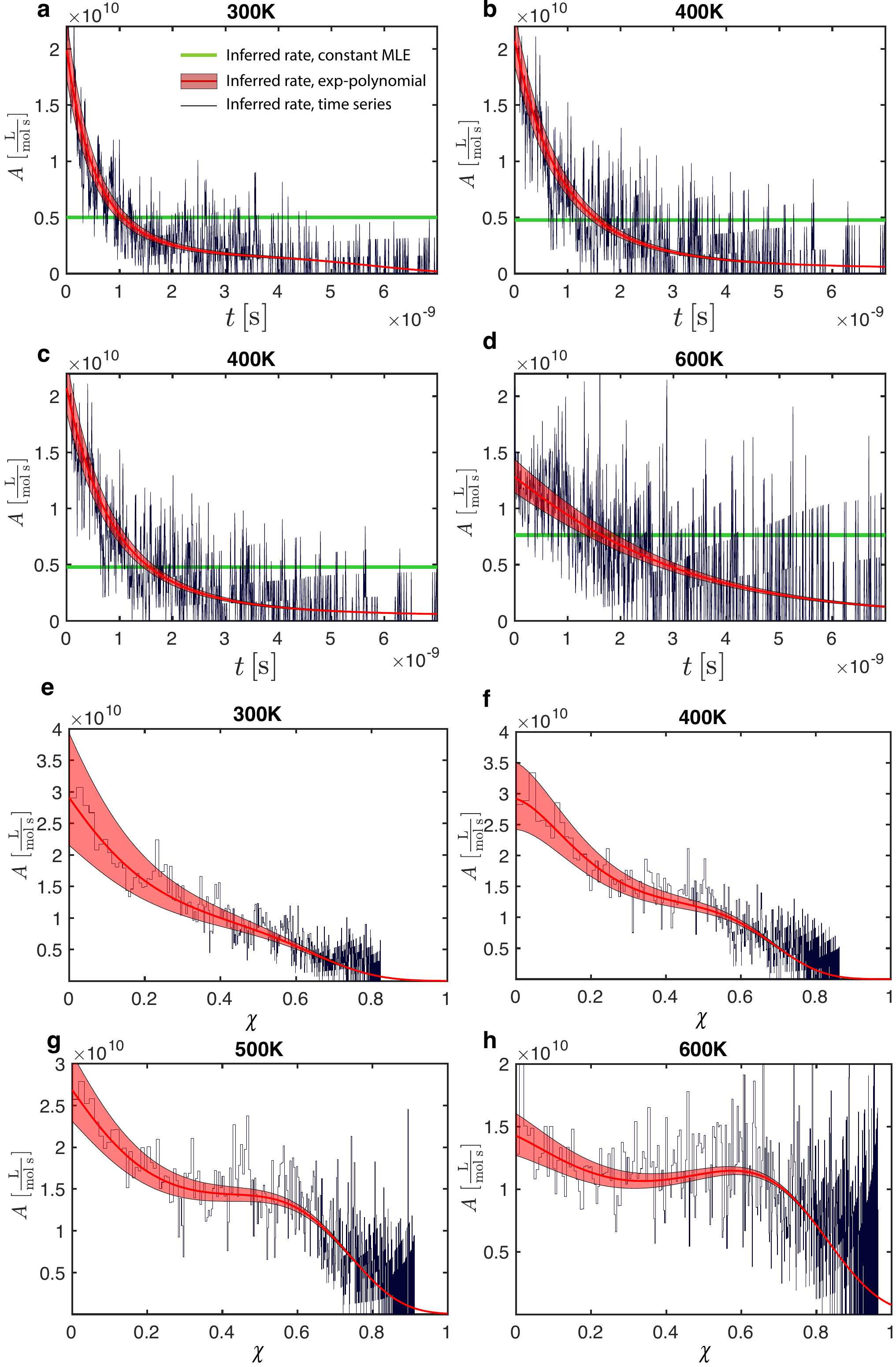}
\caption{
(a,b,c,d)  Inferred reaction rate pre-factors $A(t)$ from a single MD trajectory. Horizontal lines represent the constant estimator, equation~\eqref{eq:constMLE},  and bands  the $4^\text{rd}$ order exp-polynomial estimator, equation~\eqref{eq:MLE_EXP4}.
 Solid lines correspond to the time series estimator~\eqref{eq:seriesMLE}. The margins indicate two-standard-deviations confidence. 
(e,f,g,h) Inferred pre-factors $A(\chi)$ with time-series and exp-polynomial estimators shown.  All panels share the same legend.
}
\label{fig:rates1}
\end{center}
\end{figure}
In this section, we illustrate application of the estimators on a real world example. 
We infer the reaction rates of polymer network formation as captured by the MD simulations illustrated in Figure~\ref{fig:boxes} and show how to replace these computationally expensive MD simulations with a simple system of ODEs that are valid on arbitrary large time scales.

\paragraph*{System setup.}
Our \emph{microsystem} \cite{torres2018} is as follows: 
 2000 diacrylate molecules confined in a $7.52 \times10^{-25}\si{m^3}$ simulation box with periodic boundary conditions and integrated in time up to $10^{-8}\si{s}$ in the NPT ensemble.  Initially, 5\% of all monomers are set to be active (bearing radicals), and the activation energy of the reaction has been reduced to speed up the simulations. The true kinetic parameters can be recovered by appropriate un-biasing procedure (see Ref. \cite{torres2018} for the discussion). This microsystem is confronted with the \emph{macrosystem} that reflects the desired real world target:  $4.7\si{mol}$ of monomer units (which is of the order $10^{24}$ particles), polymerised under continuous initialisation that maintains  a steady concentration of radicals at $10^{-4}\frac{\si{mol}}{\si{L}}$  (\emph{e.g.} photo polymerisation). We investigate the rates of the two most important species: vinyl groups (V) and a radicals (R) that react via two reaction channels, respectively propagation and termination:
\begin{equation}\label{eq:eg2}
\begin{aligned}
&\ce{ V + R ->R},\\
&\ce{ R + R ->\varnothing }.
\end{aligned}
\end{equation}
 This mechanism is characterised by
$$
S=
\left(\begin{matrix}
-1&\;\;\,0\,\\
\;\;\,0& -1/2\,
\end{matrix}\right),\; \boldsymbol \nu_1=(1, 1),\; \boldsymbol \nu_2=(0, 2),
$$
which in combination with molecular dynamics data $\tilde{\boldsymbol x}_{l}$ and $\tilde{\boldsymbol z}_{l}$, provides enough information to apply the rate estimators. Since the activation energy $E_a$ has been reduced in the microsystem, we use the following decomposition of the rate:
\begin{equation}\label{eq:A(t)}
k(t)=A(t)e^{-E_a/(RT)},
\end{equation}
and perform the inference solely for pre-exponential factor $A(t)$, which is expected to be most sensitive to the interferencies from to the network formation. Here, $T$ denotes the temperature and $R$ the gas constant. 
To recover the rate coefficient $k(t)$, equation \eqref{eq:A(t)} should be supplied  with  $E_{a,1}=31.02 \si{\frac{\si{kJ}}{\si{mol}}}$ for propagation and $E_{a,2}=8.673\frac{\si{kJ}}{\si{mol}}$ for termination reactions (activation energies from the RMGpy database \footnote{RMGpy kinetic database: https://rmg.mit.edu/}).
\begin{table}[htp]
\begin{center}
\begin{tabular}{c l c l l c l}
   \hline
\hspace{0.1cm}  \vspace{0.1cm}  T[\si{K}]\hspace{0.1cm}        &
 \multicolumn{3}{c}{  \hspace{-0.1cm} Propagation, $k_1$ [$\frac{\si{mol}}{\si{L s}}$] \hspace{0.1cm}        }       &  
  \multicolumn{3}{c}{ \hspace{-0.10cm}  Termination, $k_2$ [$\frac{\si{mol}}{\si{L s}}$]\hspace{0.2cm}  }\\
   \hline
200  &  $14.55$   & $\pm$ & $0.2802$   			  &    $1.282\,{10}^6 $ &  $\pm$ & $  4.488\,{10}^5$ \\
250  &  $792.3$   & $\pm$ & $15.04$    			  &    $7.865\,{10}^6 $ &  $\pm$ & $  2.387\,{10}^6$ \\
300  &  $17733.0$   & $\pm$ & $268.7$      		  &    $2.113\,{10}^7 $ &  $\pm$ & $  5.492\,{10}^6$ \\
350  &  $97422.0$   & $\pm$ & $1385.0$    		  &    $2.494\,{10}^7 $ &  $\pm$ & $  5.964\,{10}^6$ \\
400  &  $4.276\,{10}^5$   & $\pm$ & $5301.0$  		   &    $3.106\,{10}^7 $ &  $\pm$ & $  6.898\,{10}^6$ \\
450  &  $1.682\,{10}^6$   & $\pm$ & $23599.0$  	   &    $8.03\,{10}^7 $ &  $\pm$ & $  1.890\,{10}^7$ \\
500  &  $3.644\,{10}^6$   & $\pm$ & $43900.0$      	   &    $6.258\,{10}^7 $ &  $\pm$ & $  1.336\,{10}^7$ \\
550  &  $1.031\,{10}^7$   & $\pm$ & $1.391\,{10}^5$    &    $1.237\,{10}^8 $ &  $\pm$ & $  2.871\,{10}^7$ \\
600  &  $1.64\,{10}^7$   & $\pm$ & $1.944\,{10}^5$      &    $1.515\,{10}^8 $ &  $\pm$ & $  3.577\,{10}^7$ \\
\end{tabular}
\end{center}
\caption{Inferred reaction rate parameters for HDDA polymerisation as given by the constant MLE. Confidence interval indicate two standard deviations.}
\label{tab:ratesCONST}
\end{table}%

\paragraph{Estimated Kinetic Rates.}
Table \ref{tab:ratesCONST} reports the constant rate estimations obtained with equation \eqref{eq:constMLE}.
These estimates correspond to the prefactors indicated by the horizontal lines in Figure~\ref{fig:rates1}a,b,c, and d.

According to the variance analysis given in Appendix VI, the exp-polynomial estimator was found to yield optimal estimates using $4^\text{th}$ order polynomials for the propagation reaction and $3^\text{th}$ for the termination, that is:
\begin{equation}\label{eqs:sup:MLE4}
\begin{aligned}
&k_1(\chi,T) = A_1(t)e^{-\frac{E_{a,1}}{RT}} =\\
&Ce^{- (\alpha_{1,4} \chi^4+\alpha_{1,3} \chi^3+\alpha_{1,2} \chi^2+\alpha_{1,1} \chi+\alpha_{1,0})}e^{-\frac{E_{a,1}}{RT}},
\end{aligned}
\end{equation}
and
\begin{equation}\label{eqs:sup:MLE3}
\begin{aligned}
k_2(\chi,T) = &A_2(t)e^{-\frac{E_{a,2}}{RT}} =\\
& Ce^{- (\alpha_{2,3} \chi^3+\alpha_{2,2} \chi^2+\alpha_{2,1} \chi+\alpha_{2,0})}e^{-\frac{E_{a,2}}{RT}},
\end{aligned}
\end{equation}
where the scaling constant is $C =V N_a =  452.93 \frac{ \si{L}}{\si{mol}}$.  
Instead of time $t$, we characterise the progress of the network formation by
 \begin{equation}\label{eq:p}
 \chi(t) =\frac{\#V(0)-\#V(t)}{\#V(0)}.
 \end{equation}
This quantity is also known as the bond conversion in chemistry, or occupancy probability in the theory of percolation. The coefficients are given in  Table~\ref{tab:sup:ratesMLE4}.

\def\arraystretch{1}
\begin{table*}[htp]\begin{center} \begin{tabular}{c  S[table-number-alignment = center]S[table-number-alignment = center]S[table-number-alignment = center]S[table-number-alignment = center]S[table-number-alignment = center]   | S[table-number-alignment = center] S[table-number-alignment = center] S[table-number-alignment = center] S[table-number-alignment = center] }
\hline 
T [\si{K}] & \multicolumn{5}{c}{Propagation rate} & \multicolumn{4}{c}{Termination rate}\\
\hline
 &  \hspace{0.5cm}$\alpha_{1,4}$ \hspace{0.5cm} &
      \hspace{0.5cm} $\alpha_{1,3}$  \hspace{0.5cm}&
\hspace{0.5cm}$\alpha_{1,2}$  \hspace{0.5cm}&
\hspace{0.5cm} $\alpha_{1,1}$ \hspace{0.5cm}&
\hspace{0.5cm}$\alpha_{1,0}$ \hspace{0.5cm} &
 \hspace{0.5cm} $\alpha_{2,3}$ \hspace{0.5cm} &
 \hspace{0.5cm}$\alpha_{2,2}$  \hspace{0.5cm}&
 \hspace{0.5cm}$\alpha_{2,1}$ \hspace{0.5cm}&
 \hspace{0.5cm}$\alpha_{2,0}$ \hspace{0.5cm}    \\ 
200&    78.806     &    -67.177     &    16.329     &    2.931     &    -16.479         &       163.830     &       -156.050     &       43.836     &       -16.623    \\
250&    115.780     &    -132.800   &    48.777     &    -1.957     &   -17.741      &       56.364     &       -63.904     &       27.499     &        -17.383            \\
300&    38.288     &    -38.096     &    12.209     &    2.036     &    -17.787         &        89.115     &       -98.233     &       36.041     &       -17.526    \\
350&    17.323     &    -14.024     &    2.348     &    2.447     &   -16.529       &        41.583     &       -47.910     &       19.160     &       -16.423     \\
400&    31.335     &    -36.106     &    12.000     &    0.990     &    -17.584         &        84.355     &      -103.340     &       37.832     &       -17.272    \\
450&    15.275     &    -14.874     &    3.435    &    1.506    &      -16.412             &        25.825     &       -30.223     &       13.987     &      -16.426    \\
500&    13.135     &    -11.238     &    0.775     &    1.969     &    -17.098         &        47.405     &       -66.980     &       28.765     &       -16.719      \\
550&    16.446     &    -20.363     &    7.150     &    0.065     &      -16.451       &         49.559     &       -65.727     &       25.621     &       -16.464    \\
600&    15.189     &    -16.538     &    3.135     &    1.164     &     -16.178       &         39.276     &       -54.521     &       22.272     &      -15.959  \\
\end{tabular}\end{center}\caption{The coefficients for the optimal order exp-polynomial MLEs. }\label{tab:sup:ratesMLE4}\end{table*}

\begin{figure}
\begin{center}
\includegraphics[width=\columnwidth]{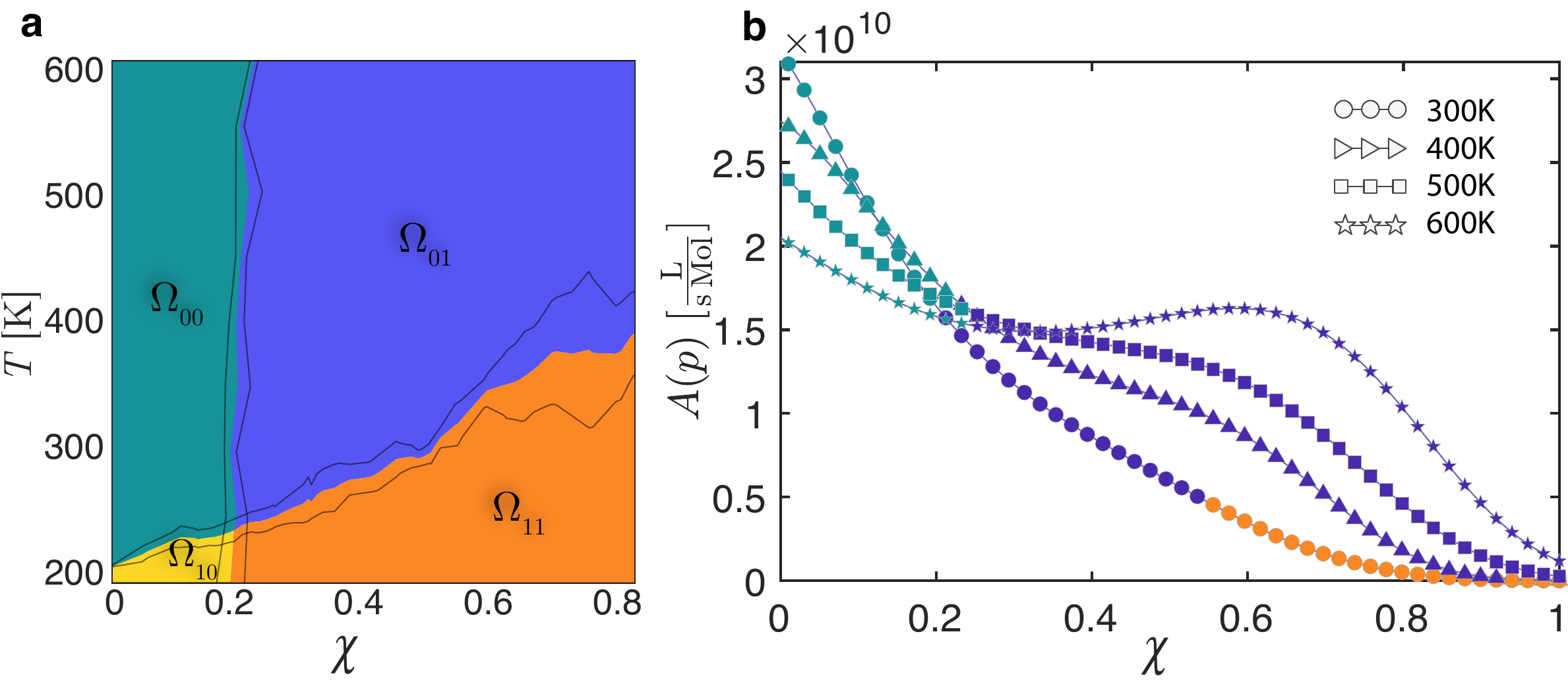}
\caption{(colour online)
{ \bf a,}  The $T-\chi$ phase space: $\Omega_{00}$ -- viscous, no network;  $\Omega_{10}$ -- glassy, no network; $\Omega_{01}$ -- rubbery, network; $\Omega_{11}$ -- glassy, network. See \cite{torres2021} for computational procedure. The solid lines mark one-standard deviation confidence interval around the domain boundaries. { \bf b,} Inferred profiles of $A(\chi)$ show that the polymerisation temperature has opposite effects on the reaction rates in different domains, $\Omega_{00},\Omega_{10}$ and $\Omega_{01},\Omega_{11}$.
 The colours code the domain of the phase space.
}
\label{fig:phase}
\end{center}
\end{figure}

\paragraph*{Nonlinear rate behaviour.}
Figure~\ref{fig:rates1} presents the inferred from single MD trajectories values of $A(t)$ and $A(\chi)$ for different temperatures of polymerisation $T$. Independently of $T$, both $A(t)$ and $A(\chi)$ strongly decrease throughout the reaction progress. 
This complex behaviour can be possibly explained by the fact that the system undergoes two phase-transitions that may not necessarily coincide: the transition from disconnected clusters to a spanning network (the percolation transition \cite{kryven2019}), and the transition from liquid/resin-like to solid/glassy state (the glass transition \cite{torres2021}).  Thus in total, we have four distinct domains in the $T-\chi$ phase space:  $\Omega_{00}$ -- viscous, no network;  $\Omega_{10}$ -- glassy, no network; $\Omega_{01}$ -- rubbery, network; $\Omega_{11}$ -- glassy, network. As shown in Figure~\ref{fig:phase}a, the partition of the phase space into these domains, indicates that the topological transition occurs around $\chi_c\approx0.2$ independently of temperature, whereas the critical value of $\chi$ for glass  transition is a function of $T$.
 
\begin{figure}
\begin{center}
\includegraphics[width=1\columnwidth]{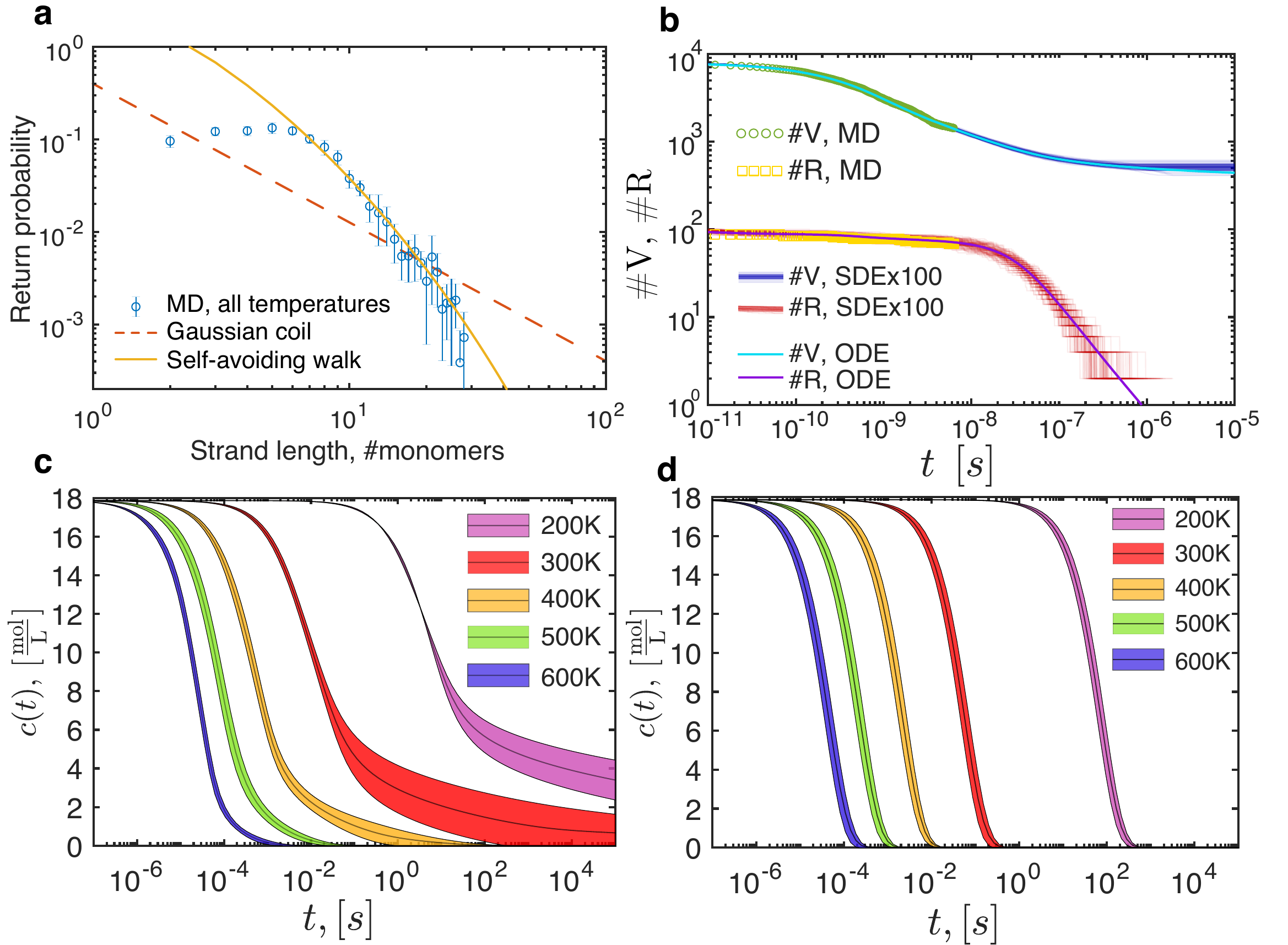}
\caption{(colour online)  
(a) The empirical probability of a network strand to close a cycle compared to Flory's self-avoiding random walk and Gaussian coil. Error bars correspond to one standard deviation.
(b,c,d) The upscaling procedure: (b) molecular simulations versus learned SDEs and ODEs with exp-polynomial coefficients,
(c)  macroscopic ODEs with exp-polynomial coefficients, (d) macroscopic ODEs with constant coefficients.}
\label{fig:micromacro}
\end{center}
\end{figure}

By colour-coding the points in the profiles of $A(\chi)$ depending to which domain they belong to, Figure \ref{fig:phase}b reveals that increasing $T$ has opposite effects on $A$ below and above the topological phase transition: increased temperature inhibits the value of pre-factor $A$ for $\chi<\chi_c$ and promotes this value for $\chi>\chi_c$.
Moreover, the collisions in a network are governed by different mechanisms than collisions in the ideal gas: shortest path between species embedded in a network becomes the most important factor that explains the collision rates, which, in turn, is independent of temperature or pressure. To emphasise the universal dependence of system's geometry on the topology we compute the return probability of the shortest path in the network when it closes a chordless cycle (a so called topological hole \cite{torres2021}). The probability that a polymer chain closes a chordelss cycle of length $n$ is typically derived from the return probability of a random walk that models the chain's geometry, however the exact definition of this random walk is a topic of  debates~\cite{rubinstein2003polymer,lang2018,wang2016,rozenfeld2005}.
As shown in Figure~\ref{fig:micromacro}a, the empirical probability that a network strand closes a cycle is universal and can be asymptotically related to Flory's expression for the self-avoiding random walk, $$p\sim n^{-3/2}e^{-\frac{3}{2}n^{-1}-\alpha n^{1/2}},$$
where the chain stiffness parameter  $\alpha=1.2$ was found by fitting. The fact that the return probability does not depend on temperature is exclusive to networks since the latter feature more geometrically constrained configurations as compared to loose chains.

\paragraph*{Upscaling.}
 The most important applied implication of the rate inference is that one can use this procedure to perform predictions with the accuracy close to that of molecular simulations but on the macroscopic scale.  Since all kinetic parameters are derived from the particle potentials, as encoded by the force field, such predictions can be almost parameter-free.
In order to perform the predictions, one models the reaction mechanism \eqref{eq:eg2} with ODEs \eqref{eq:ctox} that are supplied with the inferred expressions of $A(\chi)$, where $\chi$ is given by Eq.~\eqref{eq:p}. Figures~\ref{fig:micromacro}b,c and d illustrate this principle: Fig.~\ref{fig:micromacro}b compares MD data with the stochastic and ODE models, still in the microsystem, whereas Figs.~\ref{fig:micromacro}c and d present the upscaled results as given by the ODEs with inferred rates for the macrosystem up to $t=100\si{s}$.  Note that representing the rates as exp-polynomial functions of $\chi$ (Figure~ \ref{fig:micromacro}c) as opposed to constant rates (Figure~\ref{fig:micromacro}d) is essential to capture the kinetic slowdown that is induced by the jamming and is especially pronounced at low temperatures.

\section{Conclusion}
We propose a solution of the inverse problem to Gillespie's stochastic simulation algorithm \cite{gillespie1976}: using the empirical counts of molecular species we recover the reaction rate parameters that drive the kinetics.
From the point of view of molecular dynamics, a reaction rate is an emergent phenomenon of many reactive particles, and our method allows one to extract the effective kinetic parameters from such simulations. Assuming that the inferred parameters are scale-invariant, we show  that the results of reactive molecular simulations may be upscaled in a such a way that they become descriptive at the macroscopic scale. 

Molecular simulations of many reaction-driven macroscopic phenomena are already on the way, see for example the studies on crystallisation \cite{matsumoto2002,niu2018}, self-assembly \cite{perilla2015}, aggregation \cite{buell2010}, separation \cite{Smrek2017}, and polymerisation \cite{torres2018,sarkar2018,scolari2018}, and the concept of ordinary differential equations that learn from molecular simulations may facilitate discovery of new macroscopic laws and improving existing kinetic models for these phenomena.  As a proof of concept, we applied the method to diacrylate polymerisation to reveal an intricate phenomenological dependance of the kinetic parameters on temperature and time in this system and postulate that these dependencies are induced by the complex evolution of the underlaying network.
With this example we demonstrated that it is possible to model the transition between freely interacting spices and a dense network with ordinary differential equations having non-linear coefficients. We expect that combining such MD-informed kinetic ODEs with random graphs \cite{kryven2016emergence,kryven2018analytic,schamboeck2020coloured} may result in accurate macroscopic models that also predict network related phenomena.
 
\begin{acknowledgments}
The authors are grateful to S. Woutersen for providing critical comments at the early stage of the manuscript.  
I.K. acknowledges the support from the research programme VENI with Project No 639.071.511, and A.T-K. acknowledges support from PREDAGIO Project. Both projects were financed by the Netherlands Organisation for Scientific Research (NWO).
\end{acknowledgments}

\begingroup
\let\clearpage\relax 
\onecolumngrid 
\endgroup

\section*{Appendix I: Derivation of the stochastic rate equation}
Consider a system that consists of a single molecule undergoing a first order reaction. If the reaction firing probabilities are independent and proportional to waiting time. The probability that time $t$ passes until this molecule reacts, is given by an exponential random variable with parameter $\lambda$:
$$
\mathbb P [ t \in [\tau,\tau+\text{d} \tau]]=\lambda e^{-\lambda \tau}.
$$
We refer to this fact as $t\sim\text{Exp}[\lambda],$ also known as the ``exponential clock" \cite{del2017}.  If instead, we have $x_1=\#A$ independent molecules of the same species, the time until the first reaction firing within this set of molecules is given by: 
\begin{equation}\label{eqs:poisson1}
t\sim\inf\{\underbrace{\text{Exp}[\lambda],\dots,\text{Exp}[\lambda]}_{x_1\; \text{times}}\}\sim\text{Exp}[x_1\lambda]. 
\end{equation}
Here, we made use of the standard result about the minimum of multiple exponential random variables~\cite{del2017}.
 Since $t$ is again an exponential random variable, its expected value is given by $\mathbb E [t] = (\lambda x_1)^{-1}$, which gives the characteristic time between reaction firings. Thus, the reaction rate $r$ (the amount of substance per volume per time) is given by
 \begin{equation}\label{eqs:Arr1}
r=\frac{1}{\mathbb E [t] } \frac{1}{V N_A}=  \frac{x_1\lambda }{V N_A}=\lambda c(t) =kc(t),
\end{equation}
where the last equality derives from the fact that  $c_i = \frac{x_i}{ V N_A },$ where $N_A$ is the Avogadro's number.
Hence, equation \eqref{eqs:Arr1} settles the relationship between the stochastic rate $\lambda$ and the rate constant $k$ for  first order reactions:
\begin{equation}\label{eqs:k1}
k=\lambda.
\end{equation}
The rates of second order reactions are dependent on a coincidence of two events: 1. the two reactants collide in the correct configuration, 2. together they undergo a first order reaction. We thus have a two-stage process: 
\begin{equation}\label{eqs:abc}
\ce{ A + B <=>AB -> C },
\end{equation}
 where AB is an intermediate  that represents the species that collided but have not reacted.  According to Arrhenius theory, the first stage settles on an equilibrium: the number of AB is a constant fraction of the total number of couple combinations:
   $$\#\text{AB}=  \mathcal A x_1 x_2.$$
   Since \ce{AB -> C } is a first-order mechanism, it features the stochastic rate $\lambda'$ as given by Eq.  \eqref{eqs:Arr1}.
Consequently, one writes the time until the first reaction firing as
\begin{equation}\label{eqs:poisson2}
t\sim\text{Exp}[ \lambda' \mathcal A x_1 x_2]=\text{Exp}[ \lambda x_1 x_2], \;\lambda=\lambda'\mathcal A
\end{equation}
which, after applying similar transformations to Eq.~\eqref{eqs:Arr1}, gives the approximation for the second-order reaction rate:
$$
r = \frac{1}{\mathbb E[t]} \frac{1}{V N_A}=\frac{\lambda x_1 x_2}{V N_A} = \lambda V N_A c_1(t)c_2(t)=k c_1(t)c_2(t). 
$$
Hence, for second order reactions we have: 
\begin{equation}\label{eqs:k2}
k = \lambda V N_A.
\end{equation}
Note that if a second order reaction takes place between members of the same species, then the number of couples $\#\text{AA}=\frac{1}{2}x_1(x_1-1)$ and therefore, $k\approx\frac{1}{2}\lambda V N_A.$
More generally, if the $j^\text{th}$ reaction (of arbitrary order now) is isolated, the waiting time that passes before the reaction firing is $t\sim\text{Exp}[\lambda_j\boldsymbol x^{\boldsymbol \nu_j}]$
and, by analogy to Eq.~\eqref{eqs:poisson1}, the time until the earliest event in the case of multiple competing reactions is given by: 
$t\sim\inf_j\text{Exp}[\lambda_j\boldsymbol x^{\boldsymbol \nu_j}]\sim\text{Exp}[\sum\limits_j\lambda_j\boldsymbol x^{\boldsymbol \nu_j}].$
Moreover, the probability that this is the $j^\text{th}$ reaction is given by:
$
\mathbb P[ j ] = \frac{\lambda_j\boldsymbol x^{\boldsymbol \nu_j}}{\sum\limits_i \lambda_i\boldsymbol x^{\boldsymbol \nu_i}}.
$
When iterated over multiple time steps, the later two sampling rules yield the stochastic process \eqref{eq:walk}.

\section*{Appendix II: Constant MLE}
We consider the general setting in which the time intervals $\tau_l=t_{l}-t_{l-1}, \; l=1,\dots,L$ need not be equispaced.
Let $\lambda_j(t) = \lambda_j = \text{const}$, then the rates of the Poisson random variables from Eq.~\eqref{eq:walk} are given by
$$
  \lambda_j\boldsymbol x_l^{\boldsymbol \nu_j} \tau_l,\; l=1,\dots,L.
$$
Therefore, the probability to observe configuration $\tilde{\boldsymbol x}_{l}, \tilde{\boldsymbol z}_{l}$ on time intervals $\tau_l$ is given by: 
$$\prod\limits_{l=1}^L\prod\limits_{j=1}^M e^{-\lambda }\frac{\lambda^y}{y!}\Big|_{
\resizebox{0.09\hsize}{!}{$\begin{array}{l}
y={\boldsymbol {\tilde z}}_{j,l}\\
\lambda =  \lambda_{j}{\boldsymbol {\tilde x}}_l^{\boldsymbol \nu_j}\tau_l
\end{array}$}
},$$
and taking a logarithm of this product gives the log-likelihood of the entire ensemble of data:
\begin{equation}\label{eqs:SI:MLE_CONST}
f(\lambda_{1},\dots,\lambda_{M})=\sum\limits_{ l=1}^{L}\sum\limits_{j=1}^M 
( -\lambda  +y\ln \lambda-\ln y!)\Big|_{
\resizebox{0.09\hsize}{!}{$\begin{array}{l}
y=\boldsymbol{\tilde z}_{j,l}\\
\lambda = \lambda_{j}\boldsymbol {\tilde x}_l^{\boldsymbol \nu_j}\tau_l,
\end{array}$}
}
\end{equation}
which has the following derivatives:
$$
\frac{\partial f}{\partial \lambda_{j}}=
-\sum\limits_{ l=1}^{L}\boldsymbol {\tilde x}_l^{\boldsymbol \nu_{j}}\tau_l+\frac{1}{\lambda_{j}}\sum\limits_{ l=1}^{L}\boldsymbol {\tilde z}_{j,l}=
-L\langle\boldsymbol {\tilde x}_l^{\boldsymbol \nu_{j}}\tau_l\rangle+\frac{1}{\lambda_{j}}L\langle\boldsymbol {\tilde z}_{j,l}\rangle$$
where $\langle x_l\rangle:=\frac{1}{L}\sum\limits_{l=1}^{L}x_l$.
By equating  this derivatives to zero, one obtains expressions for $\lambda_j$:
\begin{equation}\label{eqs:SI:constMLE}
\lambda_j=\frac{\langle \boldsymbol {\tilde z}_{j,l}\rangle}{\langle\boldsymbol {\tilde x}_l^{\boldsymbol \nu_{j}}\tau_l\rangle},\;j=1,\dots,M,\end{equation}

In order to give an estimate for the variance of these parameter,  $\text{var}(\lambda_1,\dots,\lambda_n),$ we make use of the asymptotic normality property of this MLE and write:
\begin{equation}\label{eqs:SI:HESS}
\text{var}(\lambda_1,\dots,\lambda_n) =  -\left( \text{Hess} f(\lambda_1,\dots,\lambda_n)] \right)^{-1},
\end{equation}
where
$\text{Hess} f(\lambda_1,\dots,\lambda_n):=\frac{\partial^2 f}{\partial k_i \partial k_j}$ is the Hessian matrix.
Evaluating this variance estimate for Eq.~\eqref{eqs:SI:MLE_CONST} results in a diagonal covariance matrix, so that:
\begin{equation}\label{eqs:SI:var_const}
\text{var}(\lambda_j) =  \frac{\lambda_{j}^2}{L\langle \boldsymbol { \tilde z}_{j,l}\rangle}.
\end{equation}

\section*{Appendix III: Moving-average MLE}
For this estimator we require time intervals $\tau_l$  to be equispaced. Consider a modification of the previous case in which for every $l=1,\dots,L$ the parameter $\lambda(t_l)$ is calculated from a local snippet of the data $\tilde{\boldsymbol x}_{l'}, \tilde{\boldsymbol z}_{l'}$, where $l'=l-s,\dots, l+s$. Here, $s=1,2,\dots$ plays role of a regularity parameter.
  We obtain the following log-likelihood function for $\lambda_{j,l}$:
$$
\begin{aligned}
f(\lambda_{1,1},\dots,\lambda_{M,L})=&
\sum\limits_{ l=l-s}^{s+l} \sum\limits_{j=1}^M
( -\lambda  +y\ln \lambda-\ln y!)\Big|_{
\resizebox{0.15\hsize}{!}{$\begin{array}{l}
y=\boldsymbol {\tilde z}_{j,l}
\lambda = \lambda_{j,l}\boldsymbol {\tilde x}_l^{\boldsymbol \nu_j}\tau_l
\end{array}$}
}=\\
&\sum\limits_{ l=l-s}^{s+l}\sum\limits_{j=1}^M
\Big( -\lambda_{j,l}\boldsymbol {\tilde x}_l^{\boldsymbol \nu_j}\tau_l  +\boldsymbol {\tilde z}_{j,l}\ln (\lambda_{j,l})
+\boldsymbol {\tilde z}_{j,l}\ln(\boldsymbol {\tilde x}_l^{\boldsymbol \nu_j}\tau_l )-\ln( \boldsymbol {\tilde z}_{j,l}!)\Big),
\end{aligned}
$$
having derivatives:
$$
\begin{aligned}
\frac{\partial f}{\partial \lambda_{j,l}}=
-\sum\limits_{ l=l-s}^{s+l}\boldsymbol {\tilde x}_l^{\boldsymbol \nu_{j}}\tau_l+\frac{1}{\lambda_{j,l}}\sum\limits_{ l=l-s}^{s+l}\boldsymbol {\tilde z}_{j,l}=
(2s+1)(\frac{1}{\lambda_{j,l}}\langle\boldsymbol {\tilde z}_{j,l}\rangle -\langle\boldsymbol {\tilde x}_l^{\boldsymbol \nu_{j}}\tau_l\rangle),
\end{aligned}
$$
where $\langle x_l\rangle_s:=\sum\limits_{l=l-s}^{l+s}x_l$ is the moving average.
By equating  these derivatives to zero, one obtains expressions for $\lambda_{j,l}$:
\begin{equation}\label{eqs:si:MLETIME}
\lambda_{j,l}=\frac{\langle\boldsymbol{\tilde z}_{j,l}\rangle_s}{\langle\boldsymbol {\tilde x}_l^{\boldsymbol \nu_{j}}\tau_l\rangle_s}.
\end{equation}
By following an analogous derivation to the one of Eq. \eqref{eqs:SI:var_const}, one also obtains the estimate for the variance:
\begin{equation}\label{eqs:SI:var_move}
\text{var}(\lambda_{j,l}) =  \frac{\lambda_{j,l}^2}{(2s+1)\langle \boldsymbol { \tilde z}_{j,l}\rangle_s}.
\end{equation}

\section*{Appendix IV: Exponential MLE}
We consider the following ansatz:
\begin{equation}\label{eqs:SI:MLE_EXPanz}
\lambda_{j}(t) = \lambda_{j,0}e^{-\alpha_j t}.
\end{equation}
By plugging  $y=\boldsymbol{\tilde z}_{j,l}$ and $\lambda =  \lambda_{j,0}e^{-\alpha_j t_l}\boldsymbol {\tilde x}_l^{\boldsymbol \nu_j}\tau_l$ into the log-likelihood function, we obtain:
\begin{equation}
\begin{aligned}\label{eqs:SI:MLE_EXP}
f(\alpha_{1,0},&\dots,\alpha_{M,0},\alpha_1,\dots,\alpha_n)= \\
&\sum\limits_{ l=1}^L\sum\limits_{j=1}^M 
( -\lambda  +y\ln \lambda-\ln y!)= 
\sum\limits_{ l=1}^{L}\sum\limits_{j=1}^M
( -\lambda_{j,0}e^{-\alpha_j t_l}\boldsymbol {\tilde x}_l^{\boldsymbol \nu_j} \tau_l 
+\boldsymbol{\tilde z}_{j,l}\ln \lambda_{j,0} -\boldsymbol{\tilde z}_{j,l}\alpha_j t_l
- \ln \boldsymbol{\tilde z}_{j,l}!).
\end{aligned}
\end{equation}
By equating to zero the partial derivatives with respect to $ \lambda_{j,0}$, we obtain:
$$\frac{\partial f}{\partial \lambda_{j,0}}=-\sum\limits_{ l=1}^{L}e^{-\alpha_j t_l}\boldsymbol {\tilde x}_l^{\boldsymbol \nu_j}\tau_l +
\frac{1}{\lambda_{j,0}}\sum\limits_{ l=1}^{L}\boldsymbol{\tilde z}_{j,l}=0,
$$
and consequently:
\begin{equation}\label{eqs:SI:k}
\lambda_{j,0} =  
\frac{\langle \boldsymbol{\tilde z}_{j,l}\rangle}{\langle e^{-\alpha_j t_l}\boldsymbol {\tilde x}_l^{\boldsymbol \nu_j}\tau_l\rangle}.
\end{equation}
In a similar fashion, we compute the derivatives with respect to $\alpha_j$ and equate them to zero to obtain:
$$
\begin{aligned}
\frac{\partial f}{\partial \alpha_j}=&
 \lambda_{j,0}\sum\limits_{ l=1}^{L}t_le^{-\alpha_j t_l}\boldsymbol {\tilde x}_l^{\boldsymbol \nu_j} \tau_l
-\sum\limits_{ l=1}^{L}\boldsymbol{\tilde z}_{j,l} t_l=
 \lambda_{j,0}L\langle t_l e^{-\alpha_j t_l}\boldsymbol {\tilde x}_l^{\boldsymbol \nu_j} \tau_l\rangle
-L\langle \boldsymbol{\tilde z}_{j,l} t_l\rangle=0.
\end{aligned}
$$
 Plugging Eq. \eqref{eqs:SI:k} in to the latter equality gives:
$$ \frac{\langle \boldsymbol{\tilde z}_{j,l}\rangle}{\langle e^{-\alpha_j t_l}\boldsymbol {\tilde x}_l^{\boldsymbol \nu_j}\tau_l\rangle}   
\langle t_le^{-\alpha_j t_l}\boldsymbol {\tilde x}_l^{\boldsymbol \nu_j}\tau_l\rangle 
-\langle\boldsymbol{\tilde z}_{j,l} t_l\rangle=0.
$$
and since $\langle e^{-\alpha_j t_l}\boldsymbol {\tilde x}_l^{\boldsymbol \nu_j}\tau_l\rangle>0$ one can multiply by this quantity on both sides to obtain:
\begin{equation}\label{eqs:SI:omega}
\langle \left( t_l\langle \boldsymbol{\tilde z}_{j,l}\rangle   -\langle\boldsymbol{\tilde z}_{j,l} t_l\rangle \right) \boldsymbol {\tilde x}_l^{\boldsymbol \nu_j}\tau_l \omega_j^{ t_l} \rangle 
=0, \; \omega_j \in [0,1],
\end{equation}
with $\alpha_j =-\ln \omega_j$.
If each of these  transcendental equations have a unique real root $\omega_j \in [0,1]$,  the MLE \eqref{eqs:SI:MLE_EXPanz} has a minimum at  $\alpha_j$.
Equation \eqref{eqs:SI:omega} can be solved numerically by, for example, the bisection method.
As a special case, when $t_l=h l,\;l=1,2,\dots,L$ are equispaced, Eqs.~\eqref{eqs:SI:omega} become polynomial equations. For each $j$: $ \alpha_j = -\frac{1}{h}\ln y$ where
\begin{equation}\label{eqs:polynomialeq}
\sum\limits_{l=1}^L a_l y^l=1
\end{equation}
and 
$ a_l=\left(l\langle \boldsymbol{\tilde z}_{j,l}\rangle   -\langle\boldsymbol{\tilde z}_{j,l} l\rangle \right) \boldsymbol {\tilde x}_l^{\boldsymbol \nu_j}$.
This equation can be solved numerically by reformulating it  as the eigenvalue problem for  the companion matrix.

Analogously to Eq. \eqref{eqs:SI:HESS}, the  variances of $\lambda_{j,0}$ and $\alpha_j$ can be  computed form the Hessian matrices of the corresponding log-likelihood functions. These matrices are not diagonal, however, at $t=0$ we have $\lambda_{j,0} e^{-\alpha_j t}=\lambda_{j,0}$ and therefore:
 $$\text{var}(\lambda_{j}) = \text{var}(\lambda_{j,0}) =  \frac{\lambda_{j,0}^2}{L\langle \boldsymbol{\tilde z}_{j,l}\rangle},$$
In similar fashion, when $t\gg1,$ 
$\lambda_{j,0} e^{-\alpha_j t}= e^{ (\frac{1}{t}\ln \lambda_{j,0} -\alpha_j) t}\approx  e^{-\alpha_j t}$ and
   $\text{var}(\alpha_{j}) =\frac{1}{L\lambda_{j,0}\langle t_l^2 e^{-\alpha_j t_l}\boldsymbol {\tilde x}_l^{\boldsymbol \nu_j} \tau_l\rangle}.$

\section*{Appendix V: Exp-polynomial MLE}
In this estimator we assume the ansatz:
\begin{equation}\label{eqs:si:MLE4}
 \lambda_j(t)= e^{-p_j(t)},
 \end{equation}
where
$$
  p_j(t)=  \alpha_{j,0} +\alpha_{j,1}t+\alpha_{j,2}t^2+\dots+\alpha_{j,s}t^S.$$
By plugging $y=\boldsymbol{\tilde z}_{j,l}$ and $\lambda =\lambda_j(t) \boldsymbol {\tilde x}_l^{\boldsymbol \nu_j}\tau_l=e^{-p_j(t)}\boldsymbol {\tilde x}_l^{\boldsymbol \nu_j}\tau_l$ into the log-likelihood function, we obtain
$$
\begin{aligned}
f(\alpha_{1,0},&\dots,\alpha_{M,s})=\sum\limits_{ l=1}^L  \sum\limits_{j=1}^M
( -\lambda +y\ln \lambda-\ln y!) = 
L \langle- e^{-p_j(t_l)}\boldsymbol {\tilde x}_l^{\boldsymbol \nu_j}\tau_l 
- \boldsymbol{\tilde z}_{j,l}p_j(t_l)+  \boldsymbol{\tilde z}_{j,l}\ln( \boldsymbol {\tilde x}_l^{\boldsymbol \nu_j}) +  
 \boldsymbol{\tilde z}_{j,l}\ln \tau_l+ \ln (\boldsymbol{\tilde z}_{j,l}!)\rangle.
\end{aligned}
$$
Which has  derivatives $\frac{\partial f}{\partial \alpha_{j,s}}=L\langle  e^{-p_j(t_l)} t_l^s\boldsymbol {\tilde x}_l^{\boldsymbol \nu_j} \tau_l- \boldsymbol{\tilde z}_{j,l} t_l^s\rangle$.
We obtain  $M \cdot S$ equations that define $\alpha_{j,s}$ by equating these derivatives to zero: 
 $$\langle  (e^{-p_j(t_l)} \boldsymbol {\tilde x}_l^{\boldsymbol \nu_j} \tau_l- \boldsymbol{\tilde z}_{j,l} )t_l^s\rangle =0.$$
As in the preceding case, the variance analysis is performed by computing the Hessian matrix of the log-likelihood function:
$$\frac{\partial^2 f}{\partial \alpha_{j_1,s_1} \partial \alpha_{j_2,s_2}}=\begin{cases}
-L\langle  e^{-p_j(t_l)}  \boldsymbol {\tilde x}_l^{\boldsymbol \nu_j} \tau_l t_l^{s_1} t_l^{s_2}\rangle, & \text{if }  j_1=j_2\\
0 & \text{if } j_1 \neq j_2,
\end{cases}
$$
so that $\text{var}(\alpha_{j,1},\alpha_{j,2},\dots,\alpha_{j,S})=\frac{1}{L} \boldsymbol H^{-1}$ where  $$H_{k,s} = \langle  e^{-p_j(t_l)}  \boldsymbol {\tilde x}_l^{\boldsymbol \nu_j} \tau_l t_l^{k} t_l^{s}\rangle.$$ 
Moreover, this covariance matrix translates into the total variance of the rate parameter logarithm in the following way:
\begin{equation}\label{eqs:H}
\text{var} ( \ln \lambda_j(t) ) = \text{ var}\left( \sum\limits_{s=0}^S \alpha_{j,s}t^s\right) =\frac{1}{L}  \boldsymbol b^{\top}\boldsymbol H^{-1} \boldsymbol b,
\end{equation}
where $\boldsymbol b = (1,t,t^2,\dots,t^S)^\top$.

\section*{Appendix VI: Variance analysis and model selection}
\begin{figure*}[htbp]
\begin{center}
\includegraphics[width=\textwidth]{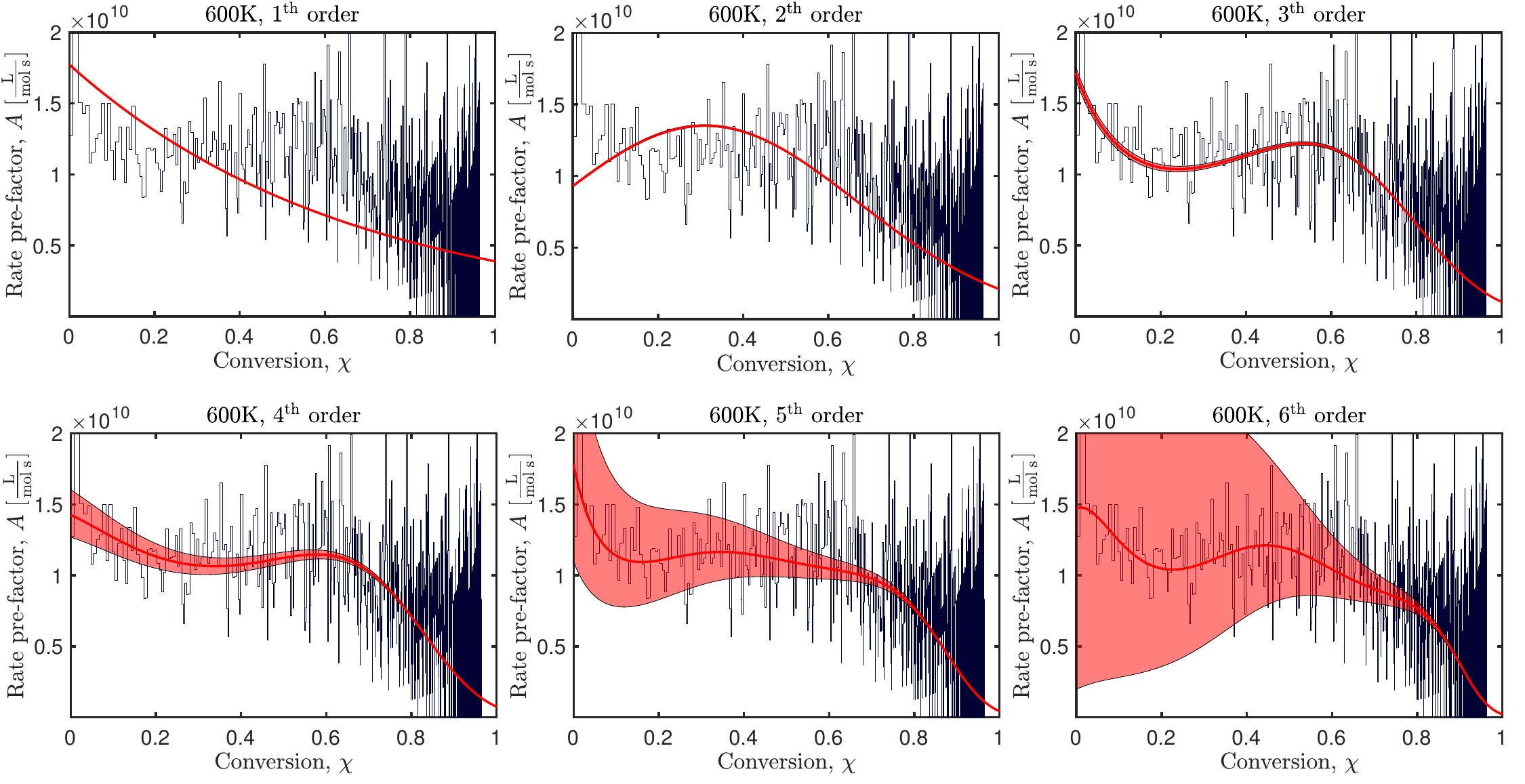}
\caption{Conversion-dependent  pre-factors as estimated with MLEs of various order (the red line plus $2\sigma$ confidence confidence intervals).
The effective time-series pre-factor is given for a reference (the black line). The optimal balance between small residual and high certainty corresponds to order 4.}
\label{SI:fig:combo4}
\end{center}
\end{figure*}
\begin{figure}[htbp]
\begin{center}
\includegraphics[width=0.73\textwidth]{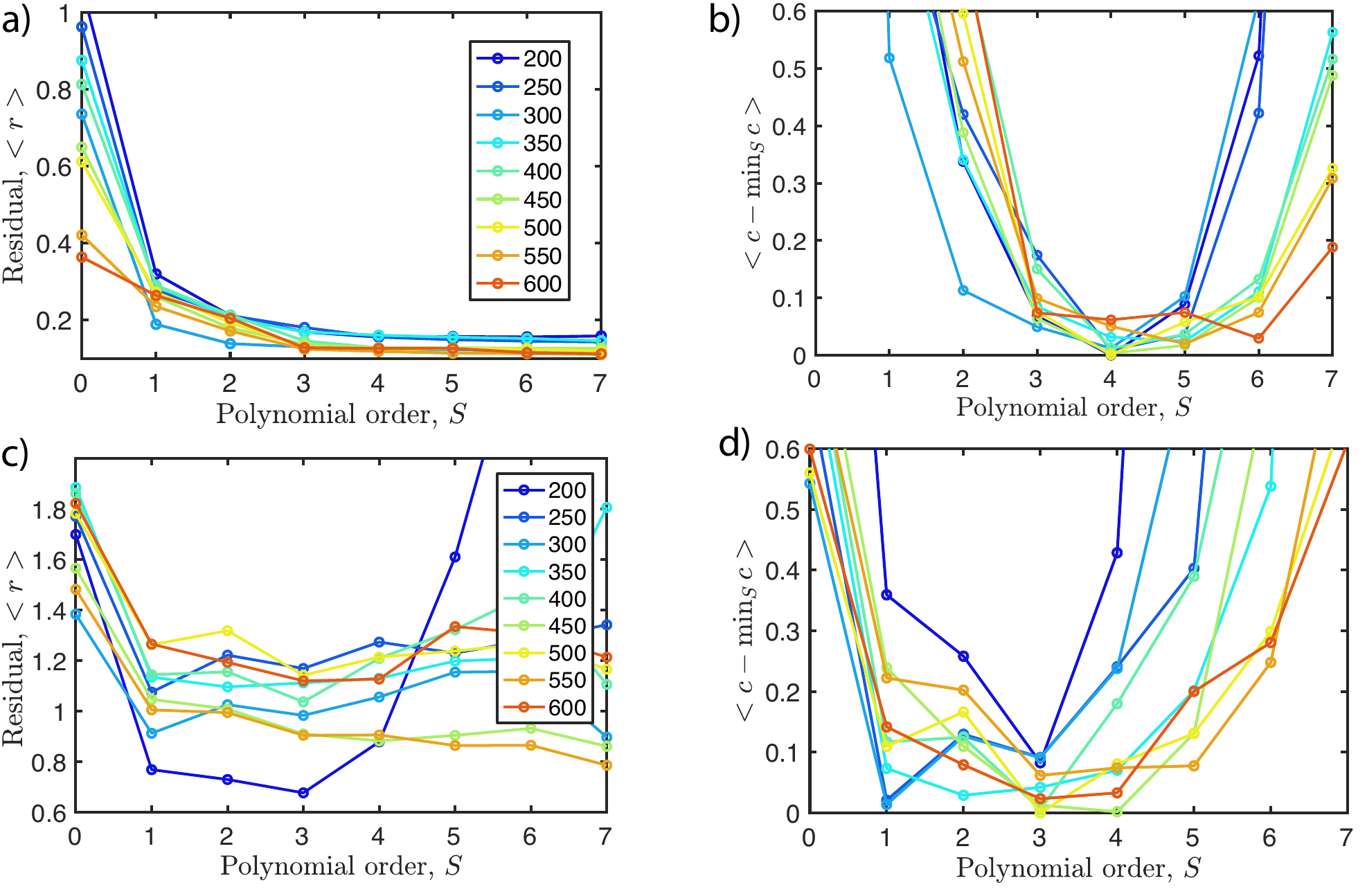}
\caption{The effect of the polynomial order $S$ of the MLE estimation of HDDA rates.
(a,b) Propagation reaction. (c,d) Termination reaction.
(a,c) The estimator residual $r$ as a function of $S$.
(b,d) The upper bound $c$ of the residual confidence interval as a function of $S$. 
The colour scheme indicates the simulation temperature. }
\label{fig:orders}
\end{center}
\end{figure}

We  consider exp-polynomial estimator~\eqref{eq:MLE_EXP4} with  conversion  $
 \chi(t) =\frac{\#V(0)-\#V(t)}{\#V(0)}
$ as the time variable.  In Figure~\ref{SI:fig:combo4} we explore how  different polynomial orders $S=0,\dots,6$ influence the inferred profiles of the rate pre-factor $A(\chi)$ and the corresponding to them confidence intervals. To quantify the quality of the exp-polynomial estimator we calculate the residual:
$
r=\int_0^1 | \ln \lambda_j(\chi) - \ln\lambda^*_j(t)| \mathrm{d} \chi,
$
where $\lambda^*_j(t)$ is given by  time-series estimator \eqref{eq:MLE_EXP4}. Generally speaking, the higher order of the polynomial the smaller are the values of $r$. Yet, this is not  the case for the variance of $r$, which has a tendency to increase with the polynomial order (the trend that can be also seen in Supplementary Figure~\ref{SI:fig:combo4}). Employing the fact that,
$\text{var}(r) = \int_0^1 \text{var}( \ln \lambda_j(\chi))  \mathrm{d}\chi^2,$
we  find the upper bound of the confidence interval to be $ c=r + 2 \sqrt{\text{var}(r)}.$
The optimal polynomial order is then  defined as the order that yields the smallest value of $c$. Figure~\ref{fig:orders}a shows that the residual indeed tends to decrease with increasing polynomial order, whereas Figure~\ref{fig:orders}b  shows that there is an optimal saddle point, $S=4$, at which the confidence interval is the smallest in the most of the MD trajectories. One can also see from Figure~\ref{fig:orders}a that the accuracy increases around 5-10 fold when we use the $4^\text{th}$ order estimator as opposed to constant one, the $0^\text{th}$ order. Similar analysis for the termination reaction reveals the optimal order of $S=3$, see Figures~\ref{fig:orders}c and d. We therefore report  the inferred  rate coefficients using  the $4^\text{th}$ polynomial for the propagation and the $3^\text{th}$ order polynomial  for the termination reaction.

\newpage
\begingroup
\let\clearpage\relax 
\twocolumngrid 
\endgroup


%

\end{document}